\documentclass[floatfix,twocolumn,aps, prl,showpacs,amsmath,amssymb]{revtex4-1}
\usepackage[latin1]{inputenc}
\usepackage[dvips]{graphicx}

\usepackage{dcolumn}
\usepackage{bm}
\usepackage{dsfont}
\usepackage{color}
\usepackage{ulem} 
\usepackage{url}
\usepackage[colorlinks=true ,citecolor=blue]{hyperref}
\usepackage{amsmath,amssymb,esint}
 \usepackage[table]{xcolor}

\definecolor{Nathanblue}{rgb}{0.,0.24,0.51}

\newcommand{\be}{\begin{equation}}
	\newcommand{\ee}{\end{equation}}
\newcommand{\bq}{\begin{eqnarray}}
	\newcommand{\eq}{\end{eqnarray}}

\begin{document}

\title{Floquet-engineering of nodal rings and nodal spheres \\and their characterization using the quantum metric}

\author{Grazia Salerno}
\affiliation{Center for Nonlinear Phenomena and Complex Systems,
	Universit\'e Libre de Bruxelles, CP 231, Campus Plaine, B-1050 Brussels, Belgium}
\author{Nathan Goldman}
\affiliation{Center for Nonlinear Phenomena and Complex Systems,
	Universit\'e Libre de Bruxelles, CP 231, Campus Plaine, B-1050 Brussels, Belgium}
\author{Giandomenico Palumbo}
\affiliation{Center for Nonlinear Phenomena and Complex Systems,
	Universit\'e Libre de Bruxelles, CP 231, Campus Plaine, B-1050 Brussels, Belgium}

\date{\today}

\begin{abstract}
Semimetals exhibiting nodal lines or nodal surfaces represent a novel class of topological states of matter. While conventional Weyl semimetals exhibit momentum-space Dirac monopoles, these more exotic semimetals can feature unusual topological defects that are analogous to extended monopoles.
In this work, we describe a scheme by which nodal rings and nodal spheres can be realized in synthetic quantum matter through well-defined periodic-driving protocols. As a central result of our work, we characterize these nodal defects through the quantum metric, which is a gauge-invariant quantity associated with the geometry of quantum states. In the case of nodal rings, where the Berry curvature and conventional topological responses are absent, we show that the quantum metric provides an observable signature for these extended topological defects. Besides, we demonstrate that quantum-metric measurements could be exploited to directly detect the topological charge associated with a nodal sphere. We discuss possible experimental implementations of Floquet nodal defects in few-level atomic systems, paving the way for the exploration of Floquet extended monopoles in quantum matter.
\end{abstract}

\maketitle

\paragraph{Introduction.} Topological states of matter play a central role in modern condensed matter physics, and they are currently under active exploration in diverse physical settings, including the solid state~\cite{Hasan-Kane_Rev}, gases of ultracold atoms~\cite{Cooper_Rev}, photonics~\cite{Ozawa_Rev} and mechanics~\cite{Huber_Rev}. These topological systems can be classified into two main categories: gapped and gapless phases.
While the former class includes topological insulators and superconductors, but also quantum Hall states~\cite{Shou-Cheng_Rev, Goerbig, Yoshioka}, the latter class concerns topological semimetals (e.g.~Weyl, Dirac and nodal-line semimetals), which have been intensively investigated in the recent years~\cite{Armitage_Rev, Kane, Morimoto-Furusaki, Balents, Fu, Dai, Chen2, Grushin, Wang, Chen, Venderbos}; besides, topological phases with nodal surfaces have also been theoretically proposed~\cite{Wang2, Timm, Sigrist, Moroz, Balatsky}. These gapless systems can display remarkable properties, such as Fermi arcs or drumhead surface states on the boundaries, and momentum-space Dirac monopoles in the bulk. 

It was proposed in Refs.~\cite{Narayan, Wang3, Gong} that two-band semimetals with a nodal line could be realized through Floquet engineering; such a scenario was recently implemented in ultracold atoms using Raman-induced spin-orbit coupling in an optical lattice~\cite{Song}. Two-band semimetals, as defined in three spatial dimensions, are fully characterized by a $\mathbb{Z}_{2}$ Berry phase around the nodal line.
However, there exist more exotic nodal line semimetals, supported by four-band models, in which a nodal ring exhibits a $\mathbb{Z}_{2}$ Berry phase and is further characterized by a $\mathbb{Z}_{2}$-monopole charge~\cite{Fu,Ahn,Tiwari}; this additional topological property substantially enhances the robustness of such nodal rings, as compared to the nodal rings of two-band models. 
More recently, these gapless phases have been related to novel three-dimensional higher-order topological insulators supporting hinge states \cite{Bernevig}.
Interestingly, the bands underlying these phases always have zero Berry curvature. This implies that optical, thermal or electric bulk responses are lacking in these settings, due to the direct relation of these observables to the Berry curvature. 
Importantly, besides the Berry curvature, there exists another gauge-invariant quantity associated with the geometry of quantum states, namely, the quantum metric \cite{Resta, Kolodrubetz}. It is known to play a role in various contexts and phenomena, including the magnetic susceptibility \cite{Piechon}, superconducting weight \cite{Peotta}, non-Abelian geometric phases \cite{Ma,Neupert}, quantum chains \cite{Gritsev}, topological insulators \cite{Palumbo2}, steady-state Hall response in driven-dissipative lattices \cite{Ozawa2}, semiclassical equations for transport \cite{Bleu,Hughes,Gao} and bulk incompressibility in fractional quantum Hall states \cite{Roy2}. It has been recently shown that the quantum metric also carries information about the topological charge ($\mathbb{Z}$-monopole) of Weyl and higher-dimensional topological semimetals \cite{Palumbo}. The quantum metric can be directly measured through dissipative responses in quantum-engineered matter, as was proposed in Ref.~\cite{Ozawa-Goldman,Ozawa_Goldman_PRR}; see also Refs.~\cite{Yu, Tan} for recent quantum-geometry measurements in NV centers and superconducting qubits. Similarly to the Berry curvature, the quantum metric can be applied, in principle, to characterize periodically-driven (Floquet) systems; see Ref.~\cite{Weitenberg}, where the integrated quantum metric (or Wannier-spread functional) was measured in a shaken optical-lattice experiment.

The aim of this Letter is twofold. First, we propose how to Floquet-engineer a nodal ring carrying a $\mathbb{Z}_{2}$-monopole charge, and a nodal sphere carrying a $\mathbb{Z}$-monopole charge. 
Second, we show that the quantum metric uniquely characterizes nodal rings and nodal spheres, and thus that it represents a direct physical signature for these unusual topological defects. A measurement of the quantum metric can therefore provide a direct signature of topological nodal objects, which could complement spectroscopic measurements~\cite{Xu}.
 
Specifically, we consider a generic four-level system with three independent parameters, supporting a 3D Dirac point. 
Inspired by Ref.~\cite{Oka}, we show that the Dirac point expands into a ring upon a suitable driving protocol; the corresponding radius is then determined by the drive parameters. In fact, the effective (Floquet) Hamiltonian~\cite{Kitagawa, GoldmanPRX, Bukov, Eckardt_Rev} then corresponds to the static model introduced in Ref.~\cite{Fu} to describe nodal line semimetals. 
We then show how a modification of the driving protocol can lead to an effective nodal sphere characterized by a $\mathbb{Z}$ topological invariant \cite{Moroz}. These inflated monopoles appear in cosmological models \cite{Vilenkin, Linde} and have been recently proposed in condensed-matter physics in free \cite{Moroz} and interacting fermionic systems \cite{Balatsky}, novel superconductors \cite{Timm} and non-Hermitian models \cite{Schnyder}.
Our work offers a simple but powerful scheme by which dynamically-induced nodal rings or nodal spheres can be created, and paves the way for the characterization of these topological defects through the quantum metric.
 
\paragraph{Floquet nodal lines.} 
We show how nodal lines can be dynamically induced through time-periodic driving in a three-dimensional parameter space. We start by considering a 3D Dirac Hamiltonian given by
\begin{eqnarray}\label{Dirac}
H_{\text{3D}}=v_D\left(\Gamma^{x}q_{x}+\Gamma^{y}q_{y}+\Gamma^{z}q_{z}\right),
\end{eqnarray}
where $v_D$ is the Dirac velocity, and where the momenta $q_{x,y,z}$ span the parameter space. 
The $4\times 4$ Dirac matrices are $\Gamma_x = \sigma_y \otimes \sigma_y$, $\Gamma_y = \mathbb{I} \otimes \sigma_x$, $\Gamma_z = -\sigma_z \otimes \sigma_y$, where $\sigma_{i}$ denote the $2\times 2$ Pauli matrices and where $\mathbb{I}$ is the $2\times 2$ identity matrix. 
This Hamiltonian is $PT$ symmetric (where $P$ is the inversion and $T$ is the time-reversal symmetry) such that the Dirac point is characterized by a $\mathbb{Z}_{2}$ monopole related to a real Berry bundle \cite{Zhao,Morimoto-Furusaki}.
The Hamiltonian in Eq.~\eqref{Dirac} is then subjected to the following time-periodic drive:
\begin{equation}\label{driving_nodal}
H_{d}(t)=A\,\Gamma^{5}\sin (\omega t)+A\,\Gamma^{x}\cos (\omega t),
\end{equation}
where $A$ is the amplitude and $\omega$ is the frequency of the drive, while the matrix $\Gamma^{5}$ is defined as $\Gamma^{5} = \sigma_x \otimes \sigma_y$.
In the following, we assume that the drive energy quantum $\hbar \omega$ is very large compared to all other energy scales in the problem, namely, $\hbar \omega \gg q^c v_D$, where the characteristic momentum $q^c$ is to be specified below Eq.~\eqref{energies}. We henceforth set $\hbar\!=\!1$, except otherwise stated. 

The total time-dependent Hamiltonian, and the corresponding effective (Floquet) Hamiltonian~\cite{Kitagawa, GoldmanPRX, Bukov, Eckardt_Rev}, are respectively defined as 
\begin{align}
&H_\text{Ring}(t) = H_{\text{3D}} + H_{d}(t), \label{Hring} \\
&H_F^{\mathrm{Ring}}=\frac{i}{T}\, \log \left[\mathcal{T} e^{-i \int_{0}^{T}dt\, H_\text{Ring}(t)}\right].\label{Floquet}
\end{align}
The Floquet Hamiltonian can be systematically expanded in powers of $1/\omega$, using standard procedures~\cite{Kitagawa, GoldmanPRX, Bukov, Eckardt_Rev}; up to first order in $1/\omega$, this reads
\begin{equation}
H_F^\mathrm{Ring} \simeq H_\text{eff}^\mathrm{Ring} \equiv H_{3D}+ i \frac{A^{2}}{\omega}\,\Gamma^{5}\Gamma^{x}.
\label{Heff}
\end{equation}
The quasi-energies of the Hamiltonian in \eqref{Heff} are
\begin{equation}\label{energies}
\varepsilon_{1,2,3,4}(\boldsymbol{q})=\pm v_D \sqrt{q_{x}^{2}+\left(\sqrt{q_{y}^{2}+q_{z}^{2}}\pm \rho(\omega)\right)^{2}},
\end{equation}
where $\rho(\omega)\equiv A^{2}/v_D\omega$. The spectrum is zero for $q_x = q_x^*=0$ and $q_{y}^{2}+q_{z}^{2}=\rho^{2}$: this second condition identifies a spectral nodal ring in the $q_y-q_z$ plane. Importantly, the ring radius $\rho=q^c$ defines the relevant characteristic momentum in this setting; consequently, one expects this Floquet-induced topological feature to be well defined in the regime $\omega \gg v_D \rho$.

\begin{figure}[t!]
	\begin{center}
		\includegraphics[width = \columnwidth]{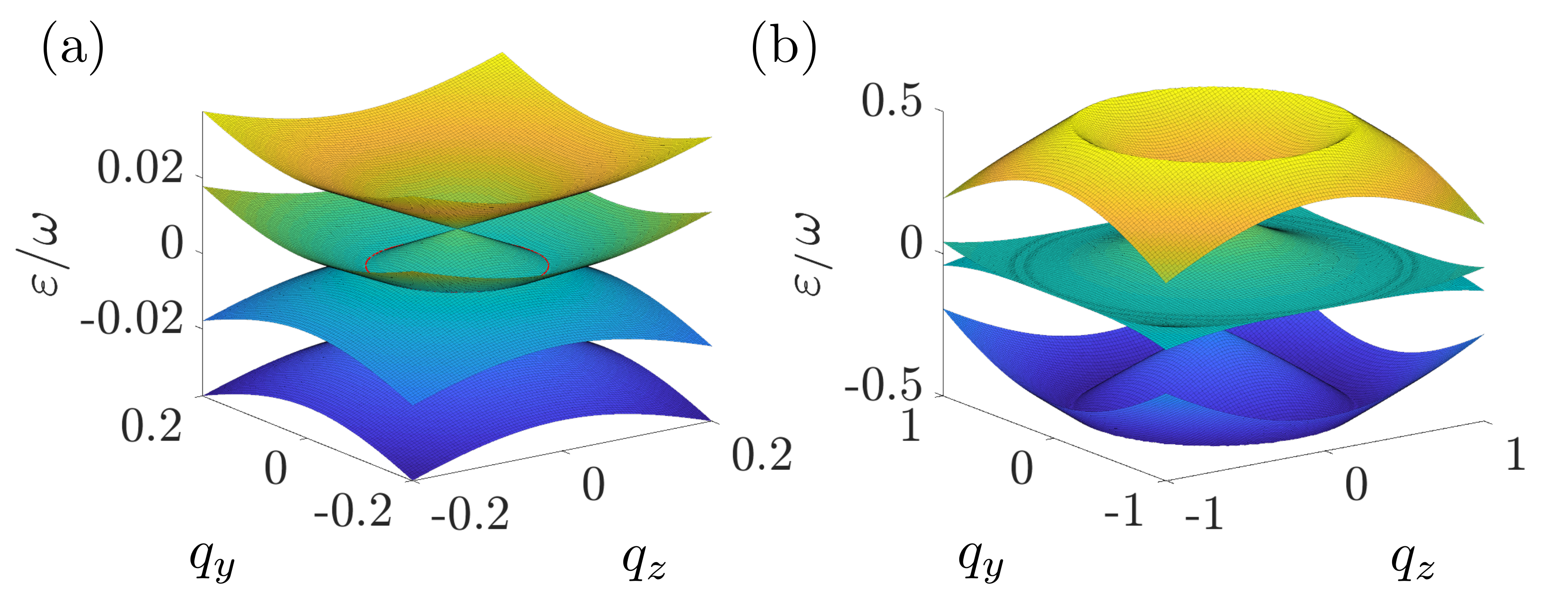}
	\end{center}
	\caption{(a) The quasi-energy band structure as computed from the full Floquet Hamiltonian in Eq.~\eqref{Floquet}, for $q_x = q_x^*=0$, and $\omega/A = 10$. These Floquet bands are in good agreement with the quasi energies found in Eq.~\eqref{energies}. The two middle bands touch in a nodal line (ring) of radius $\rho =A^{2}/v_D\omega $, shown in red. (b) The quasi-energy bands for $q_x = q_x^*=0$, and $\omega/A = 2$. The structure is more complicated, and deviates from Eq.~\eqref{energies}.}
	\label{fig:Figure1}
\end{figure} 
We calculate the quasi-energy band structure from the full Floquet Hamiltonian $H_F^\mathrm{Ring}$ defined in Eq.~\eqref{Floquet}, and show two representative spectra in Fig.~\ref{fig:Figure1}. As anticipated from the discussion above, one finds that the quasi-energies are well described by the approximate spectrum in Eq.~\eqref{energies} whenever $\omega \gg v_D \rho$, namely, for $\omega/A \gg \sqrt{2}$. This ``high-frequency" regime is illustrated in Fig.~\ref{fig:Figure1}(a), where the radius of the nodal ring (highlighted in red) agrees well with the theoretical prediction $\rho=A^{2}/v_D\omega$. Figure~\ref{fig:Figure1}(b) illustrates the ``low-frequency" regime,  $\omega \sim v_D \rho$, where higher-order corrections to the effective Hamiltonian in Eq.~\eqref{Heff} lead to important modifications of the spectrum. In particular, band-touching events at the boundary of the Floquet-Brillouin zone ($\varepsilon=\pm \omega/2$) produce an additional nodal ring; we note that this ``anomalous'' nodal defect forms a nodal sphere in the full 3D parameter space. 


While the energy bands in Eq.~\eqref{energies} are all associated with zero Berry curvature, the nodal line illustrated in Fig.~\ref{fig:Figure1}(a) carries a $\mathbb{Z}_{2}$ monopole charge~\cite{Fu}; this can be derived from the second Stiefel-Whitney invariant defined on a unitary sphere that wraps the ring~\cite{Ahn}. Besides, a direct identification of this topological invariant can be obtained by calculating a Wilson loop (associated with the touching bands) around the ring~\cite{Ahn}; we note that Wilson loops can be measured through interferometry in cold-atom setups~\cite{Bloch}. The absence of Berry curvature indicates that such topological defects cannot be signaled through standard topological responses~\cite{Armitage_Rev}. In the following, we investigate the non-vanishing quantum metric associated with the band structure; we show that this observable quantity, which can be extracted from dissipative responses~\cite{Weitenberg, Yu, Tan}, can indeed provide a direct signature for topological nodal rings.

\begin{figure}[t!]
	\begin{center}
		\includegraphics[width = \columnwidth]{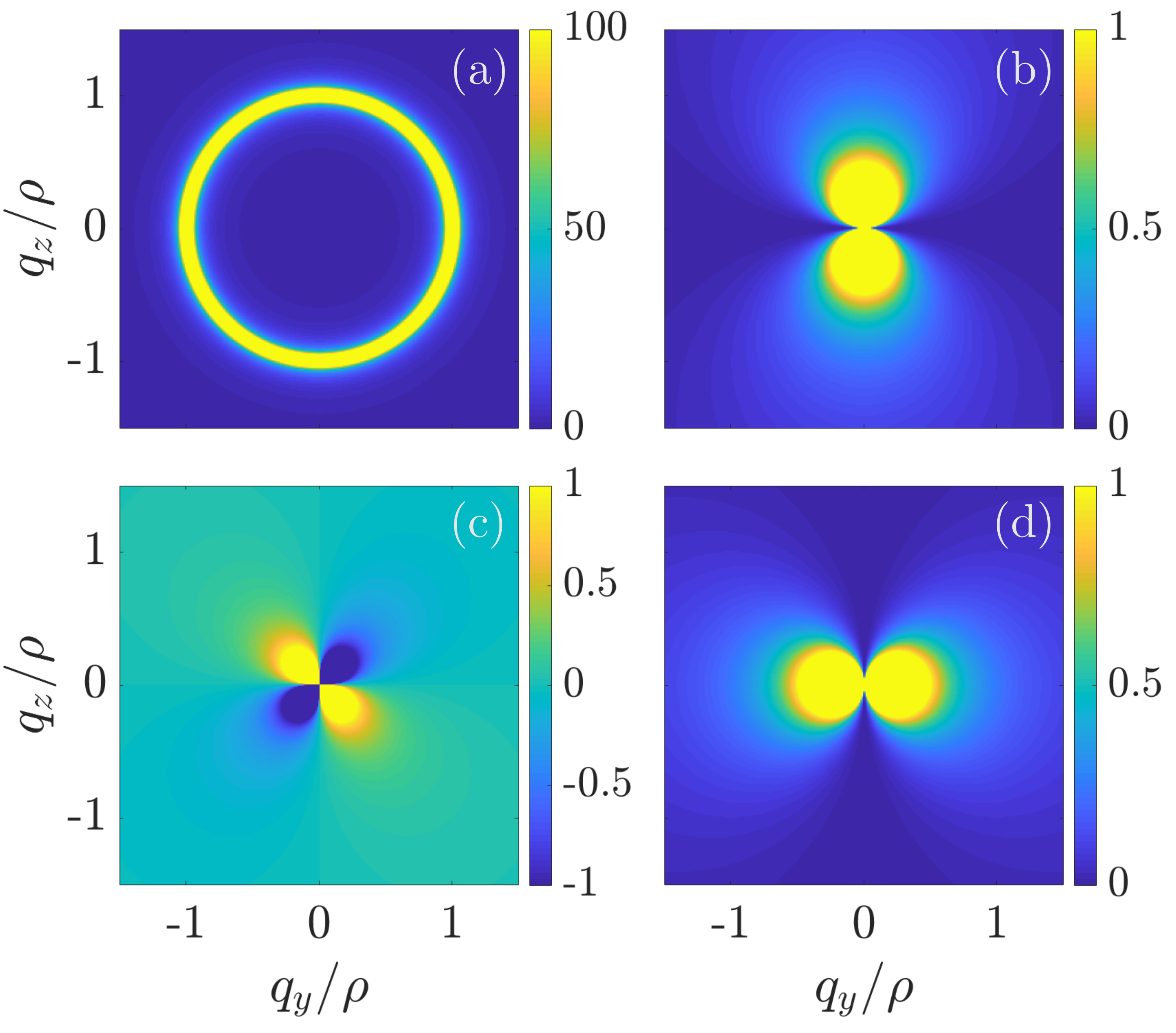}
	\end{center}
	\caption{The non-zero components of the quantum metric in Eq.~\eqref{metric} at $q_x = q_x^*$: (a) $g_{xx}$, (b) $g_{yy}$, (c) $g_{yz}$, and (d) $g_{zz}$.}
\label{fig:Figure2}
\end{figure}

In general, the three-dimensional quantum metric related to nodal lines is rather complicated. Here, we simplify the analysis by restricting the calculation of the quantum metric on the $q_{x}^*$-plane that contains the nodal ring. We find that the metric associated with the band $\varepsilon_{2}(q_y,q_z; q_{x}^{*})$, i.e.~the band containing the nodal ring, is given by
\begin{align}\label{metric}
g_{xx}|_{q_{x}^{*}}=&\frac{1}{4\left(\sqrt{q_{y}^{2}+q_{z}^{2}}-\rho(\omega)\right)^{2}},\hspace{0.2cm} 
g_{yy}|_{q_{x}^{*}}=\frac{q_{z}^{2}}{4(q_{y}^{2}+q_{z}^{2})^{2}},  \nonumber \\ 
g_{zz}|_{q_{x}^{*}}=&\frac{q_{y}^{2}}{4(q_{y}^{2}+q_{z}^{2})^{2}}, \hspace{0.2cm}
g_{yz}|_{q_{x}^{*}}=-\frac{q_{y}q_{z}}{4(q_{y}^{2}+q_{z}^{2})^{2}},\hspace{1.5cm} \nonumber \\
g_{xy}|_{q_{x}^{*}}=&\, g_{xz}|_{q_{x}^{*}}=0.
\end{align}
We note that the components $g_{yy}|_{q_{x}^{*}}$,  $g_{zz}|_{q_{x}^{*}}$ and $g_{yz}|_{q_{x}^{*}}$ coincide with the quantum metric of a 2D Dirac cone.
In this sense, the component $g_{xx}|_{q_{x}^{*}}$ is the only projected component of the quantum metric that distinguishes the nodal ring in the $q_{x}^{*}$-plane from a generic 2D Dirac cone; see Fig.~\ref{fig:Figure2}, which displays the non-zero components of the quantum metric in Eq.~\eqref{metric}. In particular, we obtain an explicit relation between the instructive component $g_{xx}|_{q_{x}^{*}}$ and the projected quasi-energies (including the nodal ring):
\begin{equation}
\varepsilon_{2,3}(\boldsymbol{q})|_{q_{x}^{*}}\equiv \pm\frac{ v_D }{2 \sqrt{g_{xx}|_{q_{x}^{*}}} }.
\end{equation}
This signature of the nodal ring is directly visible in Fig.~\ref{fig:Figure2}(a); we note that $g_{xx}$ diverges on the ring, i.e.~when $q_{y}^{2}+q_{z}^{2}=\rho^{2}$. Consequently, performing a quantum-metric tomography in the projected parameter space --  for instance, by measuring dissipative responses upon parametric modulations~\cite{Ozawa-Goldman} -- would offer a direct signature of the topological nodal ring. As a technical remark, we point out that such quantum-geometric measurements rely on preparing the system in a gapped state, i.e.~away from the ring where the component $g_{xx}|_{q_{x}^{*}}$ diverges. However, the existence of the nodal ring will still be reflected through a reconstruction of the quantum metric over the $q_{x}^{*}$-plane.

\paragraph{Floquet nodal surfaces.}
We now show how to generate a nodal sphere starting from the model in Eq.~\eqref{Dirac}. 
We consider the time-dependent Hamiltonian $H_\text{Sphere}(t) = H_{\text{3D}} + H'_{d}(t) $, in which the driving term $H'_{d}(t)$ now has the modified form
\begin{equation}\label{driving_sphere}
H_{d}'(t)=A\,\Gamma^{5}\sin (\omega t)+A\,\Gamma^{4}\cos (\omega t),
\end{equation}
with $\Gamma^{4}=-\sigma_{z} \otimes \sigma_{y}$.
Up to first-order in $1/\omega$, the Floquet Hamiltonian reads
\begin{equation}
H_F^\mathrm{Sphere} \simeq H_\text{eff}^\mathrm{Sphere} \equiv H_{\text{3D}}+ i \frac{A^{2}}{\omega}\,\Gamma^{5}\Gamma^{4}.
\label{Heff_sphere}
\end{equation}
Its quasi-energy spectrum is
\begin{equation}
\varepsilon_{1,2,3,4}(\boldsymbol{q})=\pm v_D \sqrt{q_{x}^{2}+q_{y}^{2}+q_{z}^{2}}\pm \rho(\omega),
\end{equation}
where the characteristic momentum $\rho(\omega)\!=\!A^{2}/v_D\omega$ is defined as before. In this setting, the spectrum is zero on the surface $q_{x}^{2}+q_{y}^{2}+q_{z}^{2}=\rho^{2}$, which identifies a nodal sphere embedded in  3D parameter space. We point out that such nodal spheres have been recently considered in the context of 3D nodal-surface semimetals, where the associated topological invariant was shown to be a $\mathbb{Z}$ number~\cite{Moroz}. This classification is due to the breaking of time-reversal symmetry, which allows the nodal sphere to be characterized by a first Chern number (resulting from a non-zero Berry curvature), similarly to the nodal point of standard Weyl-type semimetals. Inspired by Ref.~\cite{Palumbo}, we now discuss how this topological charge can be directly detected from the quantum metric. 

We calculate the quantum metric in an eigenstate of $H_\text{eff}^\mathrm{Sphere}$, and find the non-zero components
\begin{align}
g_{xx}= & \frac{q_{y}^2+q_{z}^2}{4\left(q_{x}^2+q_{y}^2+q_{z}^2\right)^{2}}, \qquad 
g_{yy}=\frac{q_{x}^2+q_{z}^2}{4\left(q_{x}^2+q_{y}^2+q_{z}^2\right)^{2}},  \nonumber \\ 
g_{zz}= &\frac{q_{x}^2+q_{y}^2}{4\left(q_{x}^2+q_{y}^2+q_{z}^2\right)^{2}}, \qquad
g_{xy}=-\frac{q_{x} q_{y}}{4\left(q_{x}^2+q_{y}^2+q_{z}^2\right)^{2}}, \nonumber \\
g_{xz}= &-\frac{q_{x} q_{z}}{4\left(q_{x}^2+q_{y}^2+q_{z}^2\right)^{2}},  \quad
g_{yz}=-\frac{q_{y} q_{z}}{4\left(q_{x}^2+q_{y}^2+q_{z}^2\right)^{2}}.\nonumber
 \end{align}
Interestingly, in contrast with the quantum metric associated with the nodal ring [Eq.~\eqref{metric}], we find that these components are all independent of the radius $\rho(\omega)$ of the nodal sphere. 
However, as for the Weyl-type cases discussed in Ref.~\cite{Palumbo}, we obtain a direct relation between these quantum-metric components and those of the Berry curvature $\Omega_{\mu\nu}$,
\begin{equation}
\Omega_{\mu\nu}=\epsilon_{\mu\nu}\, (2 \sqrt{\bar g})=\epsilon_{\mu\nu\lambda}\,\frac{q_{\lambda}}{2(q_{x}^{2}+q_{y}^{2}+q_{z}^{2})^{3/2}},\label{weyl_relation}
\end{equation}
where $\bar g\!=\!\det g_{\bar \mu \bar \nu}$ is the determinant of the $2\times2$ quantum metric tensor defined in the proper 2D subspace (e.g.~$\bar \mu,\bar \nu\!=\!\{q_x,q_y\}$ for the calculation of $\Omega_{xy}$). This indicates that the topological charge of the nodal sphere can be directly evaluated from the quantum metric, $Q=\frac{1}{2\pi}\int_{S^{2}} \Omega=\frac{1}{2\pi}\iint \epsilon_{\mu\nu}\, (2 \sqrt{\bar g}) dq^{\mu} dq^{\nu}$; this result highlights the fact that the nodal sphere can be interpreted as an ``inflated monopole" source of Berry curvature field~\cite{Balatsky}.

\paragraph{Experimental implementation.}
We hereby discuss possible experimental implementations of Floquet-induced nodal rings and nodal spheres in quantum matter. Inspired by a recent realization of the Yang monopole with ultracold atoms~\cite{Spielman}, we consider a four-level atomic system described by the following Hamiltonian 
\begin{equation}\label{cold-atom}
H_\text{atom} =  \begin{pmatrix}
E_{1} & \Omega_{12}(t) & 0 & \Omega_{14}(t)  \\
\Omega_{12}(t) & E_{2} & \Omega_{23}(t) & 0\\
0 & \Omega_{23}(t) & E_{3} & \Omega_{34}(t)\\
\Omega_{14}(t) & 0 & \Omega_{34}(t) & E_{4}
\end{pmatrix} ,
\end{equation}
where $\{ E_{i} \}$ are the energies of the four levels, and where $\Omega_{ij}(t) = \Omega_{ij} \cos(\omega_{ij} t + \phi_{ij})$ are the driving fields that couple the $i$-th and the $j$-th level, with amplitude $\Omega_{ij}$, frequency $\omega_{ij}$ and phase $\phi_{ij}$.
The Hamiltonian $H_{\text{3D}}$ in Eq.~\eqref{Dirac} can then be realized from $H_\text{atom}$, in the rotating wave approximation (RWA), upon choosing $\Omega_{14}=\Omega_{23}=\Omega_{B}$, $\Omega_{12}=\Omega_{34}=\Omega_{A}$,
$\omega_{12}=E_{2}-E_{1}$, $\omega_{23}=E_{3}-E_{2}$,
$\omega_{34}=E_{4}-E_{3}$, $\omega_{14}=E_{4}-E_{1}$, and $\phi_{12}=-\phi_{34} = \phi_A$, $\phi_{23} = \pi -\phi_{14}=0$; the mapping $H_{\text{3D}}\rightarrow H_\text{atom}$ is eventually obtained upon the parametrization $\Omega_B \rightarrow 2 v_D q_x$ and $\Omega_A e^{\pm i \phi_A} \rightarrow 2 v_D(q_y \pm i q_z)$. Here, we neglect any detuning from resonance ($\omega_{ij}=E_{j}-E_{i}+\delta$, with $\delta \rightarrow 0$).  

Besides, we note that the periodic driving defined in Eq.~\eqref{driving_nodal} can be designed by slowly modulating the amplitude of the Rabi frequency $\Omega_B$ according to $\Omega_B \rightarrow \Omega_B + A \cos(\omega t)$, where $\omega \ll \omega_{ij}$.
Upon applying the RWA, the total time-dependent Hamiltonian then becomes
\begin{align}\label{cold_atom_ringRWA}
&H_\text{RWA}^\text{Ring}(t) =  \frac{1}{2} \times \nonumber \\
&\begin{pmatrix}
\begin{smallmatrix}
0 & \Omega_A e^{i\phi_A} & 0 & -\Omega_B - \frac{A}{2} e^{i \omega t}  \\
\Omega_A e^{-i\phi_A} & 0 & \Omega_B +\frac{A}{2} e^{i \omega t} & 0\\
0 & \Omega_B + \frac{A}{2} e^{-i \omega t} & 0 & \Omega_A e^{-i\phi_A}\\
-\Omega_B - \frac{A}{2} e^{-i \omega t} & 0 & \Omega_A e^{i\phi_A}& 0
\end{smallmatrix} 
\end{pmatrix},
\end{align}
which is equivalent to $H_\text{Ring}(t)$ in Eq.~\eqref{Hring} upon the aforementioned parametrization. We point out that similar driven few-level systems were also studied in NV centers~\cite{Yu} and superconducting circuits \cite{Tan} in view of measuring the quantum metric. 

On the other hand, the implementation of 3D nodal spheres would require the realization of the following RWA Hamiltonian
\begin{align}\label{cold-atom_RWA_Sphere}
&H_\text{RWA}^\text{Sphere}(t) =  \frac{1}{2} \times \\
&\begin{pmatrix}
\begin{smallmatrix}
A\cos(\omega t)  & \Omega_A e^{i \phi_A} & 0 &  -\Omega_B -  i A \sin(\omega t) \\
\Omega_A e^{-i \phi_A} & - A\cos(\omega t) & \Omega_B - i A \sin(\omega t)& 0\\
0 & \Omega_B + i A \sin(\omega t) &  A\cos(\omega t)   & \Omega_A e^{-i \phi_A}\\
-\Omega_B + i A \sin(\omega t) & 0 & \Omega_A e^{i \phi_A} & - A\cos(\omega t) 
\end{smallmatrix}
\end{pmatrix} ,
\nonumber
\end{align}
The Hamiltonian in Eq.~\eqref{cold-atom_RWA_Sphere} is indeed equivalent to $H_\text{Sphere}(t)$ upon the same parametrization as above.

The quantum metric can be extracted in these models, by following the protocol of Ref.~\cite{Ozawa-Goldman}, namely, by monitoring excitation rates of initially-prepared eigenstates upon suitable parametric-modulations; see Refs.~\cite{Yu, Tan} for recent experimental implementations of this protocol in qubits, and Refs.~\cite{Weitenberg, Malpuech} for quantum-metric measurements in two-band systems.

\paragraph{Conclusions and outlooks.}
This work introduced a scheme by which nodal rings and nodal spheres can be generated in Floquet-engineered quantum systems. Besides, we have shown that the quantum metric captures the geometric and topological features of these nodal defects, as characterized by $\mathbb{Z}_{2}$ and $\mathbb{Z}$ invariants, respectively.
In this sense, our approach offers a direct method by which Floquet nodal lines and nodal surfaces can be detected through quantum-metric measurements. 
Besides direct applications in quantum-engineered systems, our work suggests several possible extensions, such as the implementation of tilted (type-II) nodal lines in three dimensions~\cite{Li} and topological nodal lines and nodal spheres in higher dimensions. We also emphasize that, in contrast with solid-state realizations of nodal lines, the quasi-energy  spectrum of Floquet systems can give rise to anomalous nodal spheres located at the boundary of the Floquet-Brillouin zone [Fig.~\ref{fig:Figure1}], which could lead to unusual correlated states upon adding inter-particle interactions~\cite{sedrakyan2015statistical}.

\paragraph{Acknowledgments.}
We are grateful to Tomoki Ozawa and Wei Chen for fruitful discussions. This work is supported by the ERC Starting Grant TopoCold, and the Fonds De La Recherche Scientifique (FRS-FNRS, Belgium).
 
\bibliography{FloquetNodal_Bibliography}

\begin{thebibliography}{64}%
\makeatletter
\providecommand \@ifxundefined [1]{%
 \@ifx{#1\undefined}
}%
\providecommand \@ifnum [1]{%
 \ifnum #1\expandafter \@firstoftwo
 \else \expandafter \@secondoftwo
 \fi
}%
\providecommand \@ifx [1]{%
 \ifx #1\expandafter \@firstoftwo
 \else \expandafter \@secondoftwo
 \fi
}%
\providecommand \natexlab [1]{#1}%
\providecommand \enquote  [1]{``#1''}%
\providecommand \bibnamefont  [1]{#1}%
\providecommand \bibfnamefont [1]{#1}%
\providecommand \citenamefont [1]{#1}%
\providecommand \href@noop [0]{\@secondoftwo}%
\providecommand \href [0]{\begingroup \@sanitize@url \@href}%
\providecommand \@href[1]{\@@startlink{#1}\@@href}%
\providecommand \@@href[1]{\endgroup#1\@@endlink}%
\providecommand \@sanitize@url [0]{\catcode `\\12\catcode `\$12\catcode
  `\&12\catcode `\#12\catcode `\^12\catcode `\_12\catcode `\%12\relax}%
\providecommand \@@startlink[1]{}%
\providecommand \@@endlink[0]{}%
\providecommand \url  [0]{\begingroup\@sanitize@url \@url }%
\providecommand \@url [1]{\endgroup\@href {#1}{\urlprefix }}%
\providecommand \urlprefix  [0]{URL }%
\providecommand \Eprint [0]{\href }%
\providecommand \doibase [0]{http://dx.doi.org/}%
\providecommand \selectlanguage [0]{\@gobble}%
\providecommand \bibinfo  [0]{\@secondoftwo}%
\providecommand \bibfield  [0]{\@secondoftwo}%
\providecommand \translation [1]{[#1]}%
\providecommand \BibitemOpen [0]{}%
\providecommand \bibitemStop [0]{}%
\providecommand \bibitemNoStop [0]{.\EOS\space}%
\providecommand \EOS [0]{\spacefactor3000\relax}%
\providecommand \BibitemShut  [1]{\csname bibitem#1\endcsname}%
\let\auto@bib@innerbib\@empty
\bibitem [{\citenamefont {Hasan}\ and\ \citenamefont
  {Kane}(2010)}]{Hasan-Kane_Rev}%
  \BibitemOpen
  \bibfield  {author} {\bibinfo {author} {\bibfnamefont {M.~Z.}\ \bibnamefont
  {Hasan}}\ and\ \bibinfo {author} {\bibfnamefont {C.~L.}\ \bibnamefont
  {Kane}},\ }\href {\doibase 10.1103/RevModPhys.82.3045} {\bibfield  {journal}
  {\bibinfo  {journal} {Rev. Mod. Phys.}\ }\textbf {\bibinfo {volume} {82}},\
  \bibinfo {pages} {3045} (\bibinfo {year} {2010})}\BibitemShut {NoStop}%
\bibitem [{\citenamefont {Cooper}\ \emph {et~al.}(2019)\citenamefont {Cooper},
  \citenamefont {Dalibard},\ and\ \citenamefont {Spielman}}]{Cooper_Rev}%
  \BibitemOpen
  \bibfield  {author} {\bibinfo {author} {\bibfnamefont {N.~R.}\ \bibnamefont
  {Cooper}}, \bibinfo {author} {\bibfnamefont {J.}~\bibnamefont {Dalibard}}, \
  and\ \bibinfo {author} {\bibfnamefont {I.~B.}\ \bibnamefont {Spielman}},\
  }\href {\doibase 10.1103/RevModPhys.91.015005} {\bibfield  {journal}
  {\bibinfo  {journal} {Rev. Mod. Phys.}\ }\textbf {\bibinfo {volume} {91}},\
  \bibinfo {pages} {015005} (\bibinfo {year} {2019})}\BibitemShut {NoStop}%
\bibitem [{\citenamefont {Ozawa}\ \emph {et~al.}(2018)\citenamefont {Ozawa},
  \citenamefont {Price}, \citenamefont {Amo}, \citenamefont {Goldman},
  \citenamefont {Hafezi}, \citenamefont {Lu}, \citenamefont {Rechtsman},
  \citenamefont {Schuster}, \citenamefont {Simon}, \citenamefont {Zilberberg},\
  and\ \citenamefont {Carusotto}}]{Ozawa_Rev}%
  \BibitemOpen
  \bibfield  {author} {\bibinfo {author} {\bibfnamefont {T.}~\bibnamefont
  {Ozawa}}, \bibinfo {author} {\bibfnamefont {H.~M.}\ \bibnamefont {Price}},
  \bibinfo {author} {\bibfnamefont {A.}~\bibnamefont {Amo}}, \bibinfo {author}
  {\bibfnamefont {N.}~\bibnamefont {Goldman}}, \bibinfo {author} {\bibfnamefont
  {M.}~\bibnamefont {Hafezi}}, \bibinfo {author} {\bibfnamefont
  {L.}~\bibnamefont {Lu}}, \bibinfo {author} {\bibfnamefont {M.}~\bibnamefont
  {Rechtsman}}, \bibinfo {author} {\bibfnamefont {D.}~\bibnamefont {Schuster}},
  \bibinfo {author} {\bibfnamefont {J.}~\bibnamefont {Simon}}, \bibinfo
  {author} {\bibfnamefont {O.}~\bibnamefont {Zilberberg}}, \ and\ \bibinfo
  {author} {\bibfnamefont {I.}~\bibnamefont {Carusotto}},\ }\href {\doibase
  10.1103/RevModPhys.91.015006} {\bibfield  {journal} {\bibinfo  {journal}
  {Rev. Mod. Phys.}\ }\textbf {\bibinfo {volume} {91}},\ \bibinfo {pages}
  {015006} (\bibinfo {year} {2018})}\BibitemShut {NoStop}%
\bibitem [{\citenamefont {Huber}(2016)}]{Huber_Rev}%
  \BibitemOpen
  \bibfield  {author} {\bibinfo {author} {\bibfnamefont {S.~D.}\ \bibnamefont
  {Huber}},\ }\href {\doibase 10.1038/nphys3801} {\bibfield  {journal}
  {\bibinfo  {journal} {Nature Physics}\ }\textbf {\bibinfo {volume} {12}},\
  \bibinfo {pages} {621} (\bibinfo {year} {2016})}\BibitemShut {NoStop}%
\bibitem [{\citenamefont {Qi}\ and\ \citenamefont
  {Zhang}(2011)}]{Shou-Cheng_Rev}%
  \BibitemOpen
  \bibfield  {author} {\bibinfo {author} {\bibfnamefont {X.-L.}\ \bibnamefont
  {Qi}}\ and\ \bibinfo {author} {\bibfnamefont {S.-C.}\ \bibnamefont {Zhang}},\
  }\href {\doibase 10.1103/RevModPhys.83.1057} {\bibfield  {journal} {\bibinfo
  {journal} {Rev. Mod. Phys.}\ }\textbf {\bibinfo {volume} {83}},\ \bibinfo
  {pages} {1057} (\bibinfo {year} {2011})}\BibitemShut {NoStop}%
\bibitem [{\citenamefont {Goerbig}(2009)}]{Goerbig}%
  \BibitemOpen
  \bibfield  {author} {\bibinfo {author} {\bibfnamefont {M.~O.}\ \bibnamefont
  {Goerbig}},\ }\href {https://arxiv.org/abs/0909.1998} {\emph {\bibinfo
  {title} {Quantum Hall Effects}}},\ \bibinfo {series} {Ultracold Gases and
  Quantum Information: Lecture Notes of the Les Houches Summer School in
  Singapore}, Vol.~\bibinfo {volume} {91}\ (\bibinfo  {publisher} {Oxford
  University Press},\ \bibinfo {year} {2009})\BibitemShut {NoStop}%
\bibitem [{\citenamefont {Yoshioka}(2002)}]{Yoshioka}%
  \BibitemOpen
  \bibfield  {author} {\bibinfo {author} {\bibfnamefont {D.}~\bibnamefont
  {Yoshioka}},\ }\href@noop {} {\emph {\bibinfo {title} {The quantum Hall
  effect}}}\ (\bibinfo  {publisher} {Springer, Berlin},\ \bibinfo {year}
  {2002})\BibitemShut {NoStop}%
\bibitem [{\citenamefont {Armitage}\ \emph {et~al.}(2018)\citenamefont
  {Armitage}, \citenamefont {Mele},\ and\ \citenamefont
  {Vishwanath}}]{Armitage_Rev}%
  \BibitemOpen
  \bibfield  {author} {\bibinfo {author} {\bibfnamefont {N.~P.}\ \bibnamefont
  {Armitage}}, \bibinfo {author} {\bibfnamefont {E.~J.}\ \bibnamefont {Mele}},
  \ and\ \bibinfo {author} {\bibfnamefont {A.}~\bibnamefont {Vishwanath}},\
  }\href {\doibase 10.1103/RevModPhys.90.015001} {\bibfield  {journal}
  {\bibinfo  {journal} {Rev. Mod. Phys.}\ }\textbf {\bibinfo {volume} {90}},\
  \bibinfo {pages} {015001} (\bibinfo {year} {2018})}\BibitemShut {NoStop}%
\bibitem [{\citenamefont {Young}\ \emph {et~al.}(2012)\citenamefont {Young},
  \citenamefont {Zaheer}, \citenamefont {Teo}, \citenamefont {Kane},
  \citenamefont {Mele},\ and\ \citenamefont {Rappe}}]{Kane}%
  \BibitemOpen
  \bibfield  {author} {\bibinfo {author} {\bibfnamefont {S.~M.}\ \bibnamefont
  {Young}}, \bibinfo {author} {\bibfnamefont {S.}~\bibnamefont {Zaheer}},
  \bibinfo {author} {\bibfnamefont {J.~C.~Y.}\ \bibnamefont {Teo}}, \bibinfo
  {author} {\bibfnamefont {C.~L.}\ \bibnamefont {Kane}}, \bibinfo {author}
  {\bibfnamefont {E.~J.}\ \bibnamefont {Mele}}, \ and\ \bibinfo {author}
  {\bibfnamefont {A.~M.}\ \bibnamefont {Rappe}},\ }\href {\doibase
  10.1103/PhysRevLett.108.140405} {\bibfield  {journal} {\bibinfo  {journal}
  {Phys. Rev. Lett.}\ }\textbf {\bibinfo {volume} {108}},\ \bibinfo {pages}
  {140405} (\bibinfo {year} {2012})}\BibitemShut {NoStop}%
\bibitem [{\citenamefont {Morimoto}\ and\ \citenamefont
  {Furusaki}(2014)}]{Morimoto-Furusaki}%
  \BibitemOpen
  \bibfield  {author} {\bibinfo {author} {\bibfnamefont {T.}~\bibnamefont
  {Morimoto}}\ and\ \bibinfo {author} {\bibfnamefont {A.}~\bibnamefont
  {Furusaki}},\ }\href {\doibase 10.1103/PhysRevB.89.235127} {\bibfield
  {journal} {\bibinfo  {journal} {Phys. Rev. B}\ }\textbf {\bibinfo {volume}
  {89}},\ \bibinfo {pages} {235127} (\bibinfo {year} {2014})}\BibitemShut
  {NoStop}%
\bibitem [{\citenamefont {Burkov}\ \emph {et~al.}(2011)\citenamefont {Burkov},
  \citenamefont {Hook},\ and\ \citenamefont {Balents}}]{Balents}%
  \BibitemOpen
  \bibfield  {author} {\bibinfo {author} {\bibfnamefont {A.~A.}\ \bibnamefont
  {Burkov}}, \bibinfo {author} {\bibfnamefont {M.~D.}\ \bibnamefont {Hook}}, \
  and\ \bibinfo {author} {\bibfnamefont {L.}~\bibnamefont {Balents}},\ }\href
  {\doibase 10.1103/PhysRevB.84.235126} {\bibfield  {journal} {\bibinfo
  {journal} {Phys. Rev. B}\ }\textbf {\bibinfo {volume} {84}},\ \bibinfo
  {pages} {235126} (\bibinfo {year} {2011})}\BibitemShut {NoStop}%
\bibitem [{\citenamefont {Fang}\ \emph {et~al.}(2015)\citenamefont {Fang},
  \citenamefont {Chen}, \citenamefont {Kee},\ and\ \citenamefont {Fu}}]{Fu}%
  \BibitemOpen
  \bibfield  {author} {\bibinfo {author} {\bibfnamefont {C.}~\bibnamefont
  {Fang}}, \bibinfo {author} {\bibfnamefont {Y.}~\bibnamefont {Chen}}, \bibinfo
  {author} {\bibfnamefont {H.-Y.}\ \bibnamefont {Kee}}, \ and\ \bibinfo
  {author} {\bibfnamefont {L.}~\bibnamefont {Fu}},\ }\href {\doibase
  10.1103/PhysRevB.92.081201} {\bibfield  {journal} {\bibinfo  {journal} {Phys.
  Rev. B}\ }\textbf {\bibinfo {volume} {92}},\ \bibinfo {pages} {081201}
  (\bibinfo {year} {2015})}\BibitemShut {NoStop}%
\bibitem [{\citenamefont {Fang}\ \emph {et~al.}(2016)\citenamefont {Fang},
  \citenamefont {Weng}, \citenamefont {Dai},\ and\ \citenamefont {Fang}}]{Dai}%
  \BibitemOpen
  \bibfield  {author} {\bibinfo {author} {\bibfnamefont {C.}~\bibnamefont
  {Fang}}, \bibinfo {author} {\bibfnamefont {H.}~\bibnamefont {Weng}}, \bibinfo
  {author} {\bibfnamefont {X.}~\bibnamefont {Dai}}, \ and\ \bibinfo {author}
  {\bibfnamefont {Z.}~\bibnamefont {Fang}},\ }\href@noop {} {\bibfield
  {journal} {\bibinfo  {journal} {Chinese Physics B}\ }\textbf {\bibinfo
  {volume} {25}},\ \bibinfo {pages} {117106} (\bibinfo {year}
  {2016})}\BibitemShut {NoStop}%
\bibitem [{\citenamefont {Chen}\ \emph {et~al.}(2017)\citenamefont {Chen},
  \citenamefont {Lu},\ and\ \citenamefont {Hou}}]{Chen2}%
  \BibitemOpen
  \bibfield  {author} {\bibinfo {author} {\bibfnamefont {W.}~\bibnamefont
  {Chen}}, \bibinfo {author} {\bibfnamefont {H.-Z.}\ \bibnamefont {Lu}}, \ and\
  \bibinfo {author} {\bibfnamefont {J.-M.}\ \bibnamefont {Hou}},\ }\href
  {\doibase 10.1103/PhysRevB.96.041102} {\bibfield  {journal} {\bibinfo
  {journal} {Phys. Rev. B}\ }\textbf {\bibinfo {volume} {96}},\ \bibinfo
  {pages} {041102} (\bibinfo {year} {2017})}\BibitemShut {NoStop}%
\bibitem [{\citenamefont {Behrends}\ \emph {et~al.}(2017)\citenamefont
  {Behrends}, \citenamefont {Rhim}, \citenamefont {Liu}, \citenamefont
  {Grushin},\ and\ \citenamefont {Bardarson}}]{Grushin}%
  \BibitemOpen
  \bibfield  {author} {\bibinfo {author} {\bibfnamefont {J.}~\bibnamefont
  {Behrends}}, \bibinfo {author} {\bibfnamefont {J.-W.}\ \bibnamefont {Rhim}},
  \bibinfo {author} {\bibfnamefont {S.}~\bibnamefont {Liu}}, \bibinfo {author}
  {\bibfnamefont {A.~G.}\ \bibnamefont {Grushin}}, \ and\ \bibinfo {author}
  {\bibfnamefont {J.~H.}\ \bibnamefont {Bardarson}},\ }\href {\doibase
  10.1103/PhysRevB.96.245101} {\bibfield  {journal} {\bibinfo  {journal} {Phys.
  Rev. B}\ }\textbf {\bibinfo {volume} {96}},\ \bibinfo {pages} {245101}
  (\bibinfo {year} {2017})}\BibitemShut {NoStop}%
\bibitem [{\citenamefont {Yan}\ \emph {et~al.}(2017)\citenamefont {Yan},
  \citenamefont {Bi}, \citenamefont {Shen}, \citenamefont {Lu}, \citenamefont
  {Zhang},\ and\ \citenamefont {Wang}}]{Wang}%
  \BibitemOpen
  \bibfield  {author} {\bibinfo {author} {\bibfnamefont {Z.}~\bibnamefont
  {Yan}}, \bibinfo {author} {\bibfnamefont {R.}~\bibnamefont {Bi}}, \bibinfo
  {author} {\bibfnamefont {H.}~\bibnamefont {Shen}}, \bibinfo {author}
  {\bibfnamefont {L.}~\bibnamefont {Lu}}, \bibinfo {author} {\bibfnamefont
  {S.-C.}\ \bibnamefont {Zhang}}, \ and\ \bibinfo {author} {\bibfnamefont
  {Z.}~\bibnamefont {Wang}},\ }\href {\doibase 10.1103/PhysRevB.96.041103}
  {\bibfield  {journal} {\bibinfo  {journal} {Phys. Rev. B}\ }\textbf {\bibinfo
  {volume} {96}},\ \bibinfo {pages} {041103} (\bibinfo {year}
  {2017})}\BibitemShut {NoStop}%
\bibitem [{\citenamefont {Chen}\ \emph {et~al.}(2018)\citenamefont {Chen},
  \citenamefont {Luo}, \citenamefont {Li},\ and\ \citenamefont
  {Zilberberg}}]{Chen}%
  \BibitemOpen
  \bibfield  {author} {\bibinfo {author} {\bibfnamefont {W.}~\bibnamefont
  {Chen}}, \bibinfo {author} {\bibfnamefont {K.}~\bibnamefont {Luo}}, \bibinfo
  {author} {\bibfnamefont {L.}~\bibnamefont {Li}}, \ and\ \bibinfo {author}
  {\bibfnamefont {O.}~\bibnamefont {Zilberberg}},\ }\href {\doibase
  10.1103/PhysRevLett.121.166802} {\bibfield  {journal} {\bibinfo  {journal}
  {Phys. Rev. Lett.}\ }\textbf {\bibinfo {volume} {121}},\ \bibinfo {pages}
  {166802} (\bibinfo {year} {2018})}\BibitemShut {NoStop}%
\bibitem [{\citenamefont {Gao}\ \emph {et~al.}(2019)\citenamefont {Gao},
  \citenamefont {Venderbos}, \citenamefont {Kim},\ and\ \citenamefont
  {Rappe}}]{Venderbos}%
  \BibitemOpen
  \bibfield  {author} {\bibinfo {author} {\bibfnamefont {H.}~\bibnamefont
  {Gao}}, \bibinfo {author} {\bibfnamefont {J.~W.~F.}\ \bibnamefont
  {Venderbos}}, \bibinfo {author} {\bibfnamefont {Y.}~\bibnamefont {Kim}}, \
  and\ \bibinfo {author} {\bibfnamefont {A.~M.}\ \bibnamefont {Rappe}},\ }\href
  {\doibase 10.1146/annurev-matsci-070218-010049} {\bibfield  {journal}
  {\bibinfo  {journal} {Annu. Rev. Mater. Res.}\ }\textbf {\bibinfo {volume}
  {49}},\ \bibinfo {pages} {153} (\bibinfo {year} {2019})}\BibitemShut
  {NoStop}%
\bibitem [{\citenamefont {Liang}\ \emph {et~al.}(2016)\citenamefont {Liang},
  \citenamefont {Zhou}, \citenamefont {Yu}, \citenamefont {Wang},\ and\
  \citenamefont {Weng}}]{Wang2}%
  \BibitemOpen
  \bibfield  {author} {\bibinfo {author} {\bibfnamefont {Q.-F.}\ \bibnamefont
  {Liang}}, \bibinfo {author} {\bibfnamefont {J.}~\bibnamefont {Zhou}},
  \bibinfo {author} {\bibfnamefont {R.}~\bibnamefont {Yu}}, \bibinfo {author}
  {\bibfnamefont {Z.}~\bibnamefont {Wang}}, \ and\ \bibinfo {author}
  {\bibfnamefont {H.}~\bibnamefont {Weng}},\ }\href {\doibase
  10.1103/PhysRevB.93.085427} {\bibfield  {journal} {\bibinfo  {journal} {Phys.
  Rev. B}\ }\textbf {\bibinfo {volume} {93}},\ \bibinfo {pages} {085427}
  (\bibinfo {year} {2016})}\BibitemShut {NoStop}%
\bibitem [{\citenamefont {Agterberg}\ \emph {et~al.}(2017)\citenamefont
  {Agterberg}, \citenamefont {Brydon},\ and\ \citenamefont {Timm}}]{Timm}%
  \BibitemOpen
  \bibfield  {author} {\bibinfo {author} {\bibfnamefont {D.~F.}\ \bibnamefont
  {Agterberg}}, \bibinfo {author} {\bibfnamefont {P.~M.~R.}\ \bibnamefont
  {Brydon}}, \ and\ \bibinfo {author} {\bibfnamefont {C.}~\bibnamefont
  {Timm}},\ }\href {\doibase 10.1103/PhysRevLett.118.127001} {\bibfield
  {journal} {\bibinfo  {journal} {Phys. Rev. Lett.}\ }\textbf {\bibinfo
  {volume} {118}},\ \bibinfo {pages} {127001} (\bibinfo {year}
  {2017})}\BibitemShut {NoStop}%
\bibitem [{\citenamefont {Bzdu\v{s}ek}\ and\ \citenamefont
  {Sigrist}(2017)}]{Sigrist}%
  \BibitemOpen
  \bibfield  {author} {\bibinfo {author} {\bibfnamefont {T.}~\bibnamefont
  {Bzdu\v{s}ek}}\ and\ \bibinfo {author} {\bibfnamefont {M.}~\bibnamefont
  {Sigrist}},\ }\href {\doibase 10.1103/PhysRevB.96.155105} {\bibfield
  {journal} {\bibinfo  {journal} {Phys. Rev. B}\ }\textbf {\bibinfo {volume}
  {96}},\ \bibinfo {pages} {155105} (\bibinfo {year} {2017})}\BibitemShut
  {NoStop}%
\bibitem [{\citenamefont {T{\"u}rker}\ and\ \citenamefont
  {Moroz}(2018)}]{Moroz}%
  \BibitemOpen
  \bibfield  {author} {\bibinfo {author} {\bibfnamefont {O.}~\bibnamefont
  {T{\"u}rker}}\ and\ \bibinfo {author} {\bibfnamefont {S.}~\bibnamefont
  {Moroz}},\ }\href {\doibase 10.1103/PhysRevB.97.075120} {\bibfield  {journal}
  {\bibinfo  {journal} {Phys. Rev. B}\ }\textbf {\bibinfo {volume} {97}},\
  \bibinfo {pages} {075120} (\bibinfo {year} {2018})}\BibitemShut {NoStop}%
\bibitem [{\citenamefont {Rostami}\ \emph {et~al.}(2018)\citenamefont
  {Rostami}, \citenamefont {Cappelluti},\ and\ \citenamefont
  {Balatsky}}]{Balatsky}%
  \BibitemOpen
  \bibfield  {author} {\bibinfo {author} {\bibfnamefont {H.}~\bibnamefont
  {Rostami}}, \bibinfo {author} {\bibfnamefont {E.}~\bibnamefont {Cappelluti}},
  \ and\ \bibinfo {author} {\bibfnamefont {A.~V.}\ \bibnamefont {Balatsky}},\
  }\href {\doibase 10.1103/PhysRevB.98.245114} {\bibfield  {journal} {\bibinfo
  {journal} {Phys. Rev. B}\ }\textbf {\bibinfo {volume} {98}},\ \bibinfo
  {pages} {245114} (\bibinfo {year} {2018})}\BibitemShut {NoStop}%
\bibitem [{\citenamefont {Narayan}(2016)}]{Narayan}%
  \BibitemOpen
  \bibfield  {author} {\bibinfo {author} {\bibfnamefont {A.}~\bibnamefont
  {Narayan}},\ }\href {\doibase 10.1103/PhysRevB.94.041409} {\bibfield
  {journal} {\bibinfo  {journal} {Phys. Rev. B}\ }\textbf {\bibinfo {volume}
  {94}},\ \bibinfo {pages} {041409} (\bibinfo {year} {2016})}\BibitemShut
  {NoStop}%
\bibitem [{\citenamefont {Yan}\ and\ \citenamefont {Wang}(2016)}]{Wang3}%
  \BibitemOpen
  \bibfield  {author} {\bibinfo {author} {\bibfnamefont {Z.}~\bibnamefont
  {Yan}}\ and\ \bibinfo {author} {\bibfnamefont {Z.}~\bibnamefont {Wang}},\
  }\href {\doibase 10.1103/PhysRevLett.117.087402} {\bibfield  {journal}
  {\bibinfo  {journal} {Phys. Rev. Lett.}\ }\textbf {\bibinfo {volume} {117}},\
  \bibinfo {pages} {087402} (\bibinfo {year} {2016})}\BibitemShut {NoStop}%
\bibitem [{\citenamefont {Li}\ \emph {et~al.}(2018)\citenamefont {Li},
  \citenamefont {Lee},\ and\ \citenamefont {Gong}}]{Gong}%
  \BibitemOpen
  \bibfield  {author} {\bibinfo {author} {\bibfnamefont {L.}~\bibnamefont
  {Li}}, \bibinfo {author} {\bibfnamefont {C.~H.}\ \bibnamefont {Lee}}, \ and\
  \bibinfo {author} {\bibfnamefont {J.}~\bibnamefont {Gong}},\ }\href {\doibase
  10.1103/PhysRevLett.121.036401} {\bibfield  {journal} {\bibinfo  {journal}
  {Phys. Rev. Lett.}\ }\textbf {\bibinfo {volume} {121}},\ \bibinfo {pages}
  {036401} (\bibinfo {year} {2018})}\BibitemShut {NoStop}%
\bibitem [{\citenamefont {Song}\ \emph {et~al.}(2019)\citenamefont {Song},
  \citenamefont {He}, \citenamefont {Niu}, \citenamefont {Zhang}, \citenamefont
  {Ren}, \citenamefont {Liu},\ and\ \citenamefont {Jo}}]{Song}%
  \BibitemOpen
  \bibfield  {author} {\bibinfo {author} {\bibfnamefont {B.}~\bibnamefont
  {Song}}, \bibinfo {author} {\bibfnamefont {C.}~\bibnamefont {He}}, \bibinfo
  {author} {\bibfnamefont {S.}~\bibnamefont {Niu}}, \bibinfo {author}
  {\bibfnamefont {L.}~\bibnamefont {Zhang}}, \bibinfo {author} {\bibfnamefont
  {Z.}~\bibnamefont {Ren}}, \bibinfo {author} {\bibfnamefont {X.-J.}\
  \bibnamefont {Liu}}, \ and\ \bibinfo {author} {\bibfnamefont {G.-B.}\
  \bibnamefont {Jo}},\ }\href {\doibase 10.1038/s41567-019-0564-y} {\bibfield
  {journal} {\bibinfo  {journal} {Nature Physics}\ }\textbf {\bibinfo {volume}
  {15}},\ \bibinfo {pages} {911} (\bibinfo {year} {2019})}\BibitemShut
  {NoStop}%
\bibitem [{\citenamefont {Ahn}\ \emph {et~al.}(2018)\citenamefont {Ahn},
  \citenamefont {Kim}, \citenamefont {Kim},\ and\ \citenamefont {Yang}}]{Ahn}%
  \BibitemOpen
  \bibfield  {author} {\bibinfo {author} {\bibfnamefont {J.}~\bibnamefont
  {Ahn}}, \bibinfo {author} {\bibfnamefont {D.}~\bibnamefont {Kim}}, \bibinfo
  {author} {\bibfnamefont {Y.}~\bibnamefont {Kim}}, \ and\ \bibinfo {author}
  {\bibfnamefont {B.-J.}\ \bibnamefont {Yang}},\ }\href {\doibase
  10.1103/PhysRevLett.121.106403} {\bibfield  {journal} {\bibinfo  {journal}
  {Phys. Rev. Lett.}\ }\textbf {\bibinfo {volume} {121}},\ \bibinfo {pages}
  {106403} (\bibinfo {year} {2018})}\BibitemShut {NoStop}%
\bibitem [{\citenamefont {Tiwari}\ and\ \citenamefont
  {Bzdusek}(2019)}]{Tiwari}%
  \BibitemOpen
  \bibfield  {author} {\bibinfo {author} {\bibfnamefont {A.}~\bibnamefont
  {Tiwari}}\ and\ \bibinfo {author} {\bibfnamefont {T.}~\bibnamefont
  {Bzdusek}},\ }\href@noop {} {\bibfield  {journal} {\bibinfo  {journal}
  {arXiv:1903.00018}\ } (\bibinfo {year} {2019})}\BibitemShut {NoStop}%
\bibitem [{\citenamefont {Wang}\ \emph {et~al.}(2019)\citenamefont {Wang},
  \citenamefont {Wieder}, \citenamefont {Li}, \citenamefont {Yan},\ and\
  \citenamefont {Bernevig}}]{Bernevig}%
  \BibitemOpen
  \bibfield  {author} {\bibinfo {author} {\bibfnamefont {Z.}~\bibnamefont
  {Wang}}, \bibinfo {author} {\bibfnamefont {B.~J.}\ \bibnamefont {Wieder}},
  \bibinfo {author} {\bibfnamefont {J.}~\bibnamefont {Li}}, \bibinfo {author}
  {\bibfnamefont {B.}~\bibnamefont {Yan}}, \ and\ \bibinfo {author}
  {\bibfnamefont {B.~A.}\ \bibnamefont {Bernevig}},\ }\href {\doibase
  10.1103/PhysRevLett.123.186401} {\bibfield  {journal} {\bibinfo  {journal}
  {Physical Review Letters}\ }\textbf {\bibinfo {volume} {123}},\ \bibinfo
  {pages} {186401} (\bibinfo {year} {2019})}\BibitemShut {NoStop}%
\bibitem [{\citenamefont {Resta}(2011)}]{Resta}%
  \BibitemOpen
  \bibfield  {author} {\bibinfo {author} {\bibfnamefont {R.}~\bibnamefont
  {Resta}},\ }\href {\doibase 10.1140/epjb/e2010-10874-4} {\bibfield  {journal}
  {\bibinfo  {journal} {The European Physical Journal B}\ }\textbf {\bibinfo
  {volume} {79}},\ \bibinfo {pages} {121} (\bibinfo {year} {2011})}\BibitemShut
  {NoStop}%
\bibitem [{\citenamefont {Kolodrubetz}\ \emph {et~al.}(2017)\citenamefont
  {Kolodrubetz}, \citenamefont {Sels}, \citenamefont {Mehta},\ and\
  \citenamefont {Polkovnikov}}]{Kolodrubetz}%
  \BibitemOpen
  \bibfield  {author} {\bibinfo {author} {\bibfnamefont {M.}~\bibnamefont
  {Kolodrubetz}}, \bibinfo {author} {\bibfnamefont {D.}~\bibnamefont {Sels}},
  \bibinfo {author} {\bibfnamefont {P.}~\bibnamefont {Mehta}}, \ and\ \bibinfo
  {author} {\bibfnamefont {A.}~\bibnamefont {Polkovnikov}},\ }\href {\doibase
  10.1016/j.physrep.2017.07.001} {\bibfield  {journal} {\bibinfo  {journal}
  {Physics Reports}\ }\textbf {\bibinfo {volume} {697}},\ \bibinfo {pages} {1}
  (\bibinfo {year} {2017})}\BibitemShut {NoStop}%
\bibitem [{\citenamefont {Pi\'echon}\ \emph {et~al.}(2016)\citenamefont
  {Pi\'echon}, \citenamefont {Raoux}, \citenamefont {Fuchs},\ and\
  \citenamefont {Montambaux}}]{Piechon}%
  \BibitemOpen
  \bibfield  {author} {\bibinfo {author} {\bibfnamefont {F.}~\bibnamefont
  {Pi\'echon}}, \bibinfo {author} {\bibfnamefont {A.}~\bibnamefont {Raoux}},
  \bibinfo {author} {\bibfnamefont {J.-N.}\ \bibnamefont {Fuchs}}, \ and\
  \bibinfo {author} {\bibfnamefont {G.}~\bibnamefont {Montambaux}},\ }\href
  {\doibase 10.1103/PhysRevB.94.134423} {\bibfield  {journal} {\bibinfo
  {journal} {Phys. Rev. B}\ }\textbf {\bibinfo {volume} {94}},\ \bibinfo
  {pages} {134423} (\bibinfo {year} {2016})}\BibitemShut {NoStop}%
\bibitem [{\citenamefont {Julku}\ \emph {et~al.}(2016)\citenamefont {Julku},
  \citenamefont {Peotta}, \citenamefont {Vanhala}, \citenamefont {Kim},\ and\
  \citenamefont {T\"orm\"a}}]{Peotta}%
  \BibitemOpen
  \bibfield  {author} {\bibinfo {author} {\bibfnamefont {A.}~\bibnamefont
  {Julku}}, \bibinfo {author} {\bibfnamefont {S.}~\bibnamefont {Peotta}},
  \bibinfo {author} {\bibfnamefont {T.~I.}\ \bibnamefont {Vanhala}}, \bibinfo
  {author} {\bibfnamefont {D.-H.}\ \bibnamefont {Kim}}, \ and\ \bibinfo
  {author} {\bibfnamefont {P.}~\bibnamefont {T\"orm\"a}},\ }\href {\doibase
  10.1103/PhysRevLett.117.045303} {\bibfield  {journal} {\bibinfo  {journal}
  {Phys. Rev. Lett.}\ }\textbf {\bibinfo {volume} {117}},\ \bibinfo {pages}
  {045303} (\bibinfo {year} {2016})}\BibitemShut {NoStop}%
\bibitem [{\citenamefont {Ma}\ \emph {et~al.}(2010)\citenamefont {Ma},
  \citenamefont {Chen}, \citenamefont {Fan},\ and\ \citenamefont {Liu}}]{Ma}%
  \BibitemOpen
  \bibfield  {author} {\bibinfo {author} {\bibfnamefont {Y.-Q.}\ \bibnamefont
  {Ma}}, \bibinfo {author} {\bibfnamefont {S.}~\bibnamefont {Chen}}, \bibinfo
  {author} {\bibfnamefont {H.}~\bibnamefont {Fan}}, \ and\ \bibinfo {author}
  {\bibfnamefont {W.-M.}\ \bibnamefont {Liu}},\ }\href {\doibase
  10.1103/PhysRevB.81.245129} {\bibfield  {journal} {\bibinfo  {journal} {Phys.
  Rev. B}\ }\textbf {\bibinfo {volume} {81}},\ \bibinfo {pages} {245129}
  (\bibinfo {year} {2010})}\BibitemShut {NoStop}%
\bibitem [{\citenamefont {Neupert}\ \emph {et~al.}(2013)\citenamefont
  {Neupert}, \citenamefont {Chamon},\ and\ \citenamefont {Mudry}}]{Neupert}%
  \BibitemOpen
  \bibfield  {author} {\bibinfo {author} {\bibfnamefont {T.}~\bibnamefont
  {Neupert}}, \bibinfo {author} {\bibfnamefont {C.}~\bibnamefont {Chamon}}, \
  and\ \bibinfo {author} {\bibfnamefont {C.}~\bibnamefont {Mudry}},\ }\href
  {\doibase 10.1103/PhysRevB.87.245103} {\bibfield  {journal} {\bibinfo
  {journal} {Phys. Rev. B}\ }\textbf {\bibinfo {volume} {87}},\ \bibinfo
  {pages} {245103} (\bibinfo {year} {2013})}\BibitemShut {NoStop}%
\bibitem [{\citenamefont {Kolodrubetz}\ \emph {et~al.}(2013)\citenamefont
  {Kolodrubetz}, \citenamefont {Gritsev},\ and\ \citenamefont
  {Polkovnikov}}]{Gritsev}%
  \BibitemOpen
  \bibfield  {author} {\bibinfo {author} {\bibfnamefont {M.}~\bibnamefont
  {Kolodrubetz}}, \bibinfo {author} {\bibfnamefont {V.}~\bibnamefont
  {Gritsev}}, \ and\ \bibinfo {author} {\bibfnamefont {A.}~\bibnamefont
  {Polkovnikov}},\ }\href {\doibase 10.1103/PhysRevB.88.064304} {\bibfield
  {journal} {\bibinfo  {journal} {Phys. Rev. B}\ }\textbf {\bibinfo {volume}
  {88}},\ \bibinfo {pages} {064304} (\bibinfo {year} {2013})}\BibitemShut
  {NoStop}%
\bibitem [{\citenamefont {Palumbo}(2018)}]{Palumbo2}%
  \BibitemOpen
  \bibfield  {author} {\bibinfo {author} {\bibfnamefont {G.}~\bibnamefont
  {Palumbo}},\ }\href {\doibase 10.1140/epjp/i2018-11856-8} {\bibfield
  {journal} {\bibinfo  {journal} {The European Physical Journal Plus}\ }\textbf
  {\bibinfo {volume} {133}},\ \bibinfo {pages} {23} (\bibinfo {year}
  {2018})}\BibitemShut {NoStop}%
\bibitem [{\citenamefont {Ozawa}(2018)}]{Ozawa2}%
  \BibitemOpen
  \bibfield  {author} {\bibinfo {author} {\bibfnamefont {T.}~\bibnamefont
  {Ozawa}},\ }\href {\doibase 10.1103/PhysRevB.97.041108} {\bibfield  {journal}
  {\bibinfo  {journal} {Phys. Rev. B}\ }\textbf {\bibinfo {volume} {97}},\
  \bibinfo {pages} {041108} (\bibinfo {year} {2018})}\BibitemShut {NoStop}%
\bibitem [{\citenamefont {Bleu}\ \emph {et~al.}(2018)\citenamefont {Bleu},
  \citenamefont {Malpuech}, \citenamefont {Gao},\ and\ \citenamefont
  {Solnyshkov}}]{Bleu}%
  \BibitemOpen
  \bibfield  {author} {\bibinfo {author} {\bibfnamefont {O.}~\bibnamefont
  {Bleu}}, \bibinfo {author} {\bibfnamefont {G.}~\bibnamefont {Malpuech}},
  \bibinfo {author} {\bibfnamefont {Y.}~\bibnamefont {Gao}}, \ and\ \bibinfo
  {author} {\bibfnamefont {D.~D.}\ \bibnamefont {Solnyshkov}},\ }\href
  {\doibase 10.1103/PhysRevLett.121.020401} {\bibfield  {journal} {\bibinfo
  {journal} {Phys. Rev. Lett.}\ }\textbf {\bibinfo {volume} {121}},\ \bibinfo
  {pages} {020401} (\bibinfo {year} {2018})}\BibitemShut {NoStop}%
\bibitem [{\citenamefont {Lapa}\ and\ \citenamefont {Hughes}(2019)}]{Hughes}%
  \BibitemOpen
  \bibfield  {author} {\bibinfo {author} {\bibfnamefont {M.~F.}\ \bibnamefont
  {Lapa}}\ and\ \bibinfo {author} {\bibfnamefont {T.~L.}\ \bibnamefont
  {Hughes}},\ }\href {\doibase 10.1103/PhysRevB.99.121111} {\bibfield
  {journal} {\bibinfo  {journal} {Phys. Rev. B}\ }\textbf {\bibinfo {volume}
  {99}},\ \bibinfo {pages} {121111} (\bibinfo {year} {2019})}\BibitemShut
  {NoStop}%
\bibitem [{\citenamefont {Gao}\ and\ \citenamefont {Xiao}(2019)}]{Gao}%
  \BibitemOpen
  \bibfield  {author} {\bibinfo {author} {\bibfnamefont {Y.}~\bibnamefont
  {Gao}}\ and\ \bibinfo {author} {\bibfnamefont {D.}~\bibnamefont {Xiao}},\
  }\href {\doibase 10.1103/PhysRevLett.122.227402} {\bibfield  {journal}
  {\bibinfo  {journal} {Phys. Rev. Lett.}\ }\textbf {\bibinfo {volume} {122}},\
  \bibinfo {pages} {227402} (\bibinfo {year} {2019})}\BibitemShut {NoStop}%
\bibitem [{\citenamefont {Parameswaran}\ \emph {et~al.}(2013)\citenamefont
  {Parameswaran}, \citenamefont {Roy},\ and\ \citenamefont {Sondhi}}]{Roy2}%
  \BibitemOpen
  \bibfield  {author} {\bibinfo {author} {\bibfnamefont {S.~A.}\ \bibnamefont
  {Parameswaran}}, \bibinfo {author} {\bibfnamefont {R.}~\bibnamefont {Roy}}, \
  and\ \bibinfo {author} {\bibfnamefont {S.~L.}\ \bibnamefont {Sondhi}},\
  }\href {\doibase https://doi.org/10.1016/j.crhy.2013.04.003} {\bibfield
  {journal} {\bibinfo  {journal} {Comptes Rendus Physique}\ }\textbf {\bibinfo
  {volume} {14}},\ \bibinfo {pages} {816} (\bibinfo {year} {2013})}\BibitemShut
  {NoStop}%
\bibitem [{\citenamefont {Palumbo}\ and\ \citenamefont
  {Goldman}(2018)}]{Palumbo}%
  \BibitemOpen
  \bibfield  {author} {\bibinfo {author} {\bibfnamefont {G.}~\bibnamefont
  {Palumbo}}\ and\ \bibinfo {author} {\bibfnamefont {N.}~\bibnamefont
  {Goldman}},\ }\href {\doibase 10.1103/PhysRevLett.121.170401} {\bibfield
  {journal} {\bibinfo  {journal} {Phys. Rev. Lett.}\ }\textbf {\bibinfo
  {volume} {121}},\ \bibinfo {pages} {170401} (\bibinfo {year}
  {2018})}\BibitemShut {NoStop}%
\bibitem [{\citenamefont {Ozawa}\ and\ \citenamefont
  {Goldman}(2018)}]{Ozawa-Goldman}%
  \BibitemOpen
  \bibfield  {author} {\bibinfo {author} {\bibfnamefont {T.}~\bibnamefont
  {Ozawa}}\ and\ \bibinfo {author} {\bibfnamefont {N.}~\bibnamefont
  {Goldman}},\ }\href {\doibase 10.1103/PhysRevB.97.201117} {\bibfield
  {journal} {\bibinfo  {journal} {Phys. Rev. B}\ }\textbf {\bibinfo {volume}
  {97}},\ \bibinfo {pages} {201117} (\bibinfo {year} {2018})}\BibitemShut
  {NoStop}%
\bibitem [{\citenamefont {Ozawa}\ and\ \citenamefont
  {Goldman}(2019)}]{Ozawa_Goldman_PRR}%
  \BibitemOpen
  \bibfield  {author} {\bibinfo {author} {\bibfnamefont {T.}~\bibnamefont
  {Ozawa}}\ and\ \bibinfo {author} {\bibfnamefont {N.}~\bibnamefont
  {Goldman}},\ }\href {\doibase 10.1103/PhysRevResearch.1.032019} {\bibfield
  {journal} {\bibinfo  {journal} {Phys. Rev. Research}\ }\textbf {\bibinfo
  {volume} {1}},\ \bibinfo {pages} {032019} (\bibinfo {year}
  {2019})}\BibitemShut {NoStop}%
\bibitem [{\citenamefont {Yu}\ \emph {et~al.}(2018)\citenamefont {Yu},
  \citenamefont {Yang}, \citenamefont {Gong}, \citenamefont {Cao},
  \citenamefont {Lu}, \citenamefont {Liu}, \citenamefont {Plenio},
  \citenamefont {Jelezko}, \citenamefont {Ozawa}, \citenamefont {Goldman},
  \citenamefont {Zhang},\ and\ \citenamefont {Cai}}]{Yu}%
  \BibitemOpen
  \bibfield  {author} {\bibinfo {author} {\bibfnamefont {M.}~\bibnamefont
  {Yu}}, \bibinfo {author} {\bibfnamefont {P.}~\bibnamefont {Yang}}, \bibinfo
  {author} {\bibfnamefont {M.}~\bibnamefont {Gong}}, \bibinfo {author}
  {\bibfnamefont {Q.}~\bibnamefont {Cao}}, \bibinfo {author} {\bibfnamefont
  {Q.}~\bibnamefont {Lu}}, \bibinfo {author} {\bibfnamefont {H.}~\bibnamefont
  {Liu}}, \bibinfo {author} {\bibfnamefont {M.~B.}\ \bibnamefont {Plenio}},
  \bibinfo {author} {\bibfnamefont {F.}~\bibnamefont {Jelezko}}, \bibinfo
  {author} {\bibfnamefont {T.}~\bibnamefont {Ozawa}}, \bibinfo {author}
  {\bibfnamefont {N.}~\bibnamefont {Goldman}}, \bibinfo {author} {\bibfnamefont
  {S.}~\bibnamefont {Zhang}}, \ and\ \bibinfo {author} {\bibfnamefont
  {J.}~\bibnamefont {Cai}},\ }\href {https://arxiv.org/abs/1811.12840}
  {\bibfield  {journal} {\bibinfo  {journal} {arXiv:1811.12840}\ } (\bibinfo
  {year} {2018})}\BibitemShut {NoStop}%
\bibitem [{\citenamefont {Tan}\ \emph {et~al.}(2019)\citenamefont {Tan},
  \citenamefont {Zhang}, \citenamefont {Yang}, \citenamefont {Chu},
  \citenamefont {Zhu}, \citenamefont {Li}, \citenamefont {Yang}, \citenamefont
  {Song}, \citenamefont {Han}, \citenamefont {Li}, \citenamefont {Dong},
  \citenamefont {Yu}, \citenamefont {Yan}, \citenamefont {Zhu},\ and\
  \citenamefont {Yu}}]{Tan}%
  \BibitemOpen
  \bibfield  {author} {\bibinfo {author} {\bibfnamefont {X.}~\bibnamefont
  {Tan}}, \bibinfo {author} {\bibfnamefont {D.-W.}\ \bibnamefont {Zhang}},
  \bibinfo {author} {\bibfnamefont {Z.}~\bibnamefont {Yang}}, \bibinfo {author}
  {\bibfnamefont {J.}~\bibnamefont {Chu}}, \bibinfo {author} {\bibfnamefont
  {Y.-Q.}\ \bibnamefont {Zhu}}, \bibinfo {author} {\bibfnamefont
  {D.}~\bibnamefont {Li}}, \bibinfo {author} {\bibfnamefont {X.}~\bibnamefont
  {Yang}}, \bibinfo {author} {\bibfnamefont {S.}~\bibnamefont {Song}}, \bibinfo
  {author} {\bibfnamefont {Z.}~\bibnamefont {Han}}, \bibinfo {author}
  {\bibfnamefont {Z.}~\bibnamefont {Li}}, \bibinfo {author} {\bibfnamefont
  {Y.}~\bibnamefont {Dong}}, \bibinfo {author} {\bibfnamefont {H.-F.}\
  \bibnamefont {Yu}}, \bibinfo {author} {\bibfnamefont {H.}~\bibnamefont
  {Yan}}, \bibinfo {author} {\bibfnamefont {S.-L.}\ \bibnamefont {Zhu}}, \ and\
  \bibinfo {author} {\bibfnamefont {Y.}~\bibnamefont {Yu}},\ }\href {\doibase
  10.1103/PhysRevLett.122.210401} {\bibfield  {journal} {\bibinfo  {journal}
  {Phys. Rev. Lett.}\ }\textbf {\bibinfo {volume} {122}},\ \bibinfo {pages}
  {210401} (\bibinfo {year} {2019})}\BibitemShut {NoStop}%
\bibitem [{\citenamefont {Asteria}\ \emph {et~al.}(2019)\citenamefont
  {Asteria}, \citenamefont {Tran}, \citenamefont {Ozawa}, \citenamefont
  {Tarnowski}, \citenamefont {Rem}, \citenamefont {Fl{\"a}schner},
  \citenamefont {Sengstock}, \citenamefont {Goldman},\ and\ \citenamefont
  {Weitenberg}}]{Weitenberg}%
  \BibitemOpen
  \bibfield  {author} {\bibinfo {author} {\bibfnamefont {L.}~\bibnamefont
  {Asteria}}, \bibinfo {author} {\bibfnamefont {D.~T.}\ \bibnamefont {Tran}},
  \bibinfo {author} {\bibfnamefont {T.}~\bibnamefont {Ozawa}}, \bibinfo
  {author} {\bibfnamefont {M.}~\bibnamefont {Tarnowski}}, \bibinfo {author}
  {\bibfnamefont {B.~S.}\ \bibnamefont {Rem}}, \bibinfo {author} {\bibfnamefont
  {N.}~\bibnamefont {Fl{\"a}schner}}, \bibinfo {author} {\bibfnamefont
  {K.}~\bibnamefont {Sengstock}}, \bibinfo {author} {\bibfnamefont
  {N.}~\bibnamefont {Goldman}}, \ and\ \bibinfo {author} {\bibfnamefont
  {C.}~\bibnamefont {Weitenberg}},\ }\href {\doibase 10.1038/s41567-019-0417-8}
  {\bibfield  {journal} {\bibinfo  {journal} {Nat. Phys.}\ }\textbf {\bibinfo
  {volume} {15}},\ \bibinfo {pages} {449} (\bibinfo {year} {2019})}\BibitemShut
  {NoStop}%
\bibitem [{\citenamefont {Xu}(2019)}]{Xu}%
  \BibitemOpen
  \bibfield  {author} {\bibinfo {author} {\bibfnamefont {Y.}~\bibnamefont
  {Xu}},\ }\href {\doibase 10.1007/s11467-019-0896-1} {\bibfield  {journal}
  {\bibinfo  {journal} {Frontiers of Physics}\ }\textbf {\bibinfo {volume}
  {14}},\ \bibinfo {pages} {43402} (\bibinfo {year} {2019})}\BibitemShut
  {NoStop}%
\bibitem [{\citenamefont {Ebihara}\ \emph {et~al.}(2016)\citenamefont
  {Ebihara}, \citenamefont {Fukushima},\ and\ \citenamefont {Oka}}]{Oka}%
  \BibitemOpen
  \bibfield  {author} {\bibinfo {author} {\bibfnamefont {S.}~\bibnamefont
  {Ebihara}}, \bibinfo {author} {\bibfnamefont {K.}~\bibnamefont {Fukushima}},
  \ and\ \bibinfo {author} {\bibfnamefont {T.}~\bibnamefont {Oka}},\ }\href
  {\doibase 10.1103/PhysRevB.93.155107} {\bibfield  {journal} {\bibinfo
  {journal} {Phys. Rev. B}\ }\textbf {\bibinfo {volume} {93}},\ \bibinfo
  {pages} {155107} (\bibinfo {year} {2016})}\BibitemShut {NoStop}%
\bibitem [{\citenamefont {Kitagawa}\ \emph {et~al.}(2010)\citenamefont
  {Kitagawa}, \citenamefont {Berg}, \citenamefont {Rudner},\ and\ \citenamefont
  {Demler}}]{Kitagawa}%
  \BibitemOpen
  \bibfield  {author} {\bibinfo {author} {\bibfnamefont {T.}~\bibnamefont
  {Kitagawa}}, \bibinfo {author} {\bibfnamefont {E.}~\bibnamefont {Berg}},
  \bibinfo {author} {\bibfnamefont {M.}~\bibnamefont {Rudner}}, \ and\ \bibinfo
  {author} {\bibfnamefont {E.}~\bibnamefont {Demler}},\ }\href {\doibase
  10.1103/PhysRevB.82.235114} {\bibfield  {journal} {\bibinfo  {journal} {Phys.
  Rev. B}\ }\textbf {\bibinfo {volume} {82}},\ \bibinfo {pages} {235114}
  (\bibinfo {year} {2010})}\BibitemShut {NoStop}%
\bibitem [{\citenamefont {Goldman}\ and\ \citenamefont
  {Dalibard}(2014)}]{GoldmanPRX}%
  \BibitemOpen
  \bibfield  {author} {\bibinfo {author} {\bibfnamefont {N.}~\bibnamefont
  {Goldman}}\ and\ \bibinfo {author} {\bibfnamefont {J.}~\bibnamefont
  {Dalibard}},\ }\href {\doibase 10.1103/PhysRevX.4.031027} {\bibfield
  {journal} {\bibinfo  {journal} {Phys. Rev. X}\ }\textbf {\bibinfo {volume}
  {4}},\ \bibinfo {pages} {031027} (\bibinfo {year} {2014})}\BibitemShut
  {NoStop}%
\bibitem [{\citenamefont {Bukov}\ \emph {et~al.}(2015)\citenamefont {Bukov},
  \citenamefont {D'Alessio},\ and\ \citenamefont {Polkovnikov}}]{Bukov}%
  \BibitemOpen
  \bibfield  {author} {\bibinfo {author} {\bibfnamefont {M.}~\bibnamefont
  {Bukov}}, \bibinfo {author} {\bibfnamefont {L.}~\bibnamefont {D'Alessio}}, \
  and\ \bibinfo {author} {\bibfnamefont {A.}~\bibnamefont {Polkovnikov}},\
  }\href {\doibase 10.1080/00018732.2015.1055918} {\bibfield  {journal}
  {\bibinfo  {journal} {Advances in Physics}\ }\textbf {\bibinfo {volume}
  {64}},\ \bibinfo {pages} {139} (\bibinfo {year} {2015})}\BibitemShut
  {NoStop}%
\bibitem [{\citenamefont {Eckardt}(2017)}]{Eckardt_Rev}%
  \BibitemOpen
  \bibfield  {author} {\bibinfo {author} {\bibfnamefont {A.}~\bibnamefont
  {Eckardt}},\ }\href {\doibase 10.1103/RevModPhys.89.011004} {\bibfield
  {journal} {\bibinfo  {journal} {Rev. Mod. Phys.}\ }\textbf {\bibinfo {volume}
  {89}},\ \bibinfo {pages} {011004} (\bibinfo {year} {2017})}\BibitemShut
  {NoStop}%
\bibitem [{\citenamefont {Vilenkin}(1994)}]{Vilenkin}%
  \BibitemOpen
  \bibfield  {author} {\bibinfo {author} {\bibfnamefont {A.}~\bibnamefont
  {Vilenkin}},\ }\href {\doibase 10.1103/PhysRevLett.72.3137} {\bibfield
  {journal} {\bibinfo  {journal} {Phys. Rev. Lett.}\ }\textbf {\bibinfo
  {volume} {72}},\ \bibinfo {pages} {3137} (\bibinfo {year}
  {1994})}\BibitemShut {NoStop}%
\bibitem [{\citenamefont {Linde}(1994)}]{Linde}%
  \BibitemOpen
  \bibfield  {author} {\bibinfo {author} {\bibfnamefont {A.}~\bibnamefont
  {Linde}},\ }\href {\doibase https://doi.org/10.1016/0370-2693(94)90719-6}
  {\bibfield  {journal} {\bibinfo  {journal} {Physics Letters B}\ }\textbf
  {\bibinfo {volume} {327}},\ \bibinfo {pages} {208 } (\bibinfo {year}
  {1994})}\BibitemShut {NoStop}%
\bibitem [{\citenamefont {Rui}\ \emph {et~al.}()\citenamefont {Rui},
  \citenamefont {Hirschmann},\ and\ \citenamefont {Schnyder}}]{Schnyder}%
  \BibitemOpen
  \bibfield  {author} {\bibinfo {author} {\bibfnamefont {W.~B.}\ \bibnamefont
  {Rui}}, \bibinfo {author} {\bibfnamefont {M.~M.}\ \bibnamefont {Hirschmann}},
  \ and\ \bibinfo {author} {\bibfnamefont {A.~P.}\ \bibnamefont {Schnyder}},\
  }\href@noop {} {\bibinfo  {journal} {arXiv:1907.10417}\ }\BibitemShut
  {NoStop}%
\bibitem [{\citenamefont {Zhao}\ and\ \citenamefont {Lu}(2017)}]{Zhao}%
  \BibitemOpen
\bibfield  {journal} {  }\bibfield  {author} {\bibinfo {author} {\bibfnamefont
  {Y.~X.}\ \bibnamefont {Zhao}}\ and\ \bibinfo {author} {\bibfnamefont
  {Y.}~\bibnamefont {Lu}},\ }\href {\doibase 10.1103/PhysRevLett.118.056401}
  {\bibfield  {journal} {\bibinfo  {journal} {Phys. Rev. Lett.}\ }\textbf
  {\bibinfo {volume} {118}},\ \bibinfo {pages} {056401} (\bibinfo {year}
  {2017})}\BibitemShut {NoStop}%
\bibitem [{\citenamefont {Li}\ \emph {et~al.}(2016)\citenamefont {Li},
  \citenamefont {Duca}, \citenamefont {Reitter}, \citenamefont {Grusdt},
  \citenamefont {Demler}, \citenamefont {Endres}, \citenamefont
  {Schleier-Smith}, \citenamefont {Bloch},\ and\ \citenamefont
  {Schneider}}]{Bloch}%
  \BibitemOpen
  \bibfield  {author} {\bibinfo {author} {\bibfnamefont {T.}~\bibnamefont
  {Li}}, \bibinfo {author} {\bibfnamefont {L.}~\bibnamefont {Duca}}, \bibinfo
  {author} {\bibfnamefont {M.}~\bibnamefont {Reitter}}, \bibinfo {author}
  {\bibfnamefont {F.}~\bibnamefont {Grusdt}}, \bibinfo {author} {\bibfnamefont
  {E.}~\bibnamefont {Demler}}, \bibinfo {author} {\bibfnamefont
  {M.}~\bibnamefont {Endres}}, \bibinfo {author} {\bibfnamefont
  {M.}~\bibnamefont {Schleier-Smith}}, \bibinfo {author} {\bibfnamefont
  {I.}~\bibnamefont {Bloch}}, \ and\ \bibinfo {author} {\bibfnamefont
  {U.}~\bibnamefont {Schneider}},\ }\href {\doibase 10.1126/science.aad5812}
  {\bibfield  {journal} {\bibinfo  {journal} {Science}\ }\textbf {\bibinfo
  {volume} {352}},\ \bibinfo {pages} {1094} (\bibinfo {year}
  {2016})}\BibitemShut {NoStop}%
\bibitem [{\citenamefont {Sugawa}\ \emph {et~al.}(2018)\citenamefont {Sugawa},
  \citenamefont {Salces-Carcoba}, \citenamefont {Perry}, \citenamefont {Yue},\
  and\ \citenamefont {Spielman}}]{Spielman}%
  \BibitemOpen
  \bibfield  {author} {\bibinfo {author} {\bibfnamefont {S.}~\bibnamefont
  {Sugawa}}, \bibinfo {author} {\bibfnamefont {F.}~\bibnamefont
  {Salces-Carcoba}}, \bibinfo {author} {\bibfnamefont {A.~R.}\ \bibnamefont
  {Perry}}, \bibinfo {author} {\bibfnamefont {Y.}~\bibnamefont {Yue}}, \ and\
  \bibinfo {author} {\bibfnamefont {I.~B.}\ \bibnamefont {Spielman}},\ }\href
  {\doibase 10.1126/science.aam9031} {\bibfield  {journal} {\bibinfo  {journal}
  {Science}\ }\textbf {\bibinfo {volume} {360}},\ \bibinfo {pages} {1429}
  (\bibinfo {year} {2018})}\BibitemShut {NoStop}%
\bibitem [{\citenamefont {Gianfrate}\ \emph {et~al.}(2019)\citenamefont
  {Gianfrate}, \citenamefont {Bleu}, \citenamefont {Dominici}, \citenamefont
  {Ardizzone}, \citenamefont {Giorgi}, \citenamefont {Ballarini}, \citenamefont
  {West}, \citenamefont {Pfeiffer}, \citenamefont {Solnyshkov}, \citenamefont
  {Sanvitto},\ and\ \citenamefont {Malpuech}}]{Malpuech}%
  \BibitemOpen
  \bibfield  {author} {\bibinfo {author} {\bibfnamefont {A.}~\bibnamefont
  {Gianfrate}}, \bibinfo {author} {\bibfnamefont {O.}~\bibnamefont {Bleu}},
  \bibinfo {author} {\bibfnamefont {L.}~\bibnamefont {Dominici}}, \bibinfo
  {author} {\bibfnamefont {V.}~\bibnamefont {Ardizzone}}, \bibinfo {author}
  {\bibfnamefont {M.~D.}\ \bibnamefont {Giorgi}}, \bibinfo {author}
  {\bibfnamefont {D.}~\bibnamefont {Ballarini}}, \bibinfo {author}
  {\bibfnamefont {K.}~\bibnamefont {West}}, \bibinfo {author} {\bibfnamefont
  {L.~N.}\ \bibnamefont {Pfeiffer}}, \bibinfo {author} {\bibfnamefont {D.~D.}\
  \bibnamefont {Solnyshkov}}, \bibinfo {author} {\bibfnamefont
  {D.}~\bibnamefont {Sanvitto}}, \ and\ \bibinfo {author} {\bibfnamefont
  {G.}~\bibnamefont {Malpuech}},\ }\href {https://arxiv.org/abs/1901.03219}
  {\bibfield  {journal} {\bibinfo  {journal} {arXiv:1901:03219}\ } (\bibinfo
  {year} {2019})}\BibitemShut {NoStop}%
\bibitem [{\citenamefont {Li}\ \emph {et~al.}(2017)\citenamefont {Li},
  \citenamefont {Yu}, \citenamefont {Liu}, \citenamefont {Guan}, \citenamefont
  {Wang}, \citenamefont {Zhang}, \citenamefont {Yao},\ and\ \citenamefont
  {Yang}}]{Li}%
  \BibitemOpen
  \bibfield  {author} {\bibinfo {author} {\bibfnamefont {S.}~\bibnamefont
  {Li}}, \bibinfo {author} {\bibfnamefont {Z.-M.}\ \bibnamefont {Yu}}, \bibinfo
  {author} {\bibfnamefont {Y.}~\bibnamefont {Liu}}, \bibinfo {author}
  {\bibfnamefont {S.}~\bibnamefont {Guan}}, \bibinfo {author} {\bibfnamefont
  {S.-S.}\ \bibnamefont {Wang}}, \bibinfo {author} {\bibfnamefont
  {X.}~\bibnamefont {Zhang}}, \bibinfo {author} {\bibfnamefont
  {Y.}~\bibnamefont {Yao}}, \ and\ \bibinfo {author} {\bibfnamefont {S.~A.}\
  \bibnamefont {Yang}},\ }\href {\doibase 10.1103/PhysRevB.96.081106}
  {\bibfield  {journal} {\bibinfo  {journal} {Phys. Rev. B}\ }\textbf {\bibinfo
  {volume} {96}},\ \bibinfo {pages} {081106} (\bibinfo {year}
  {2017})}\BibitemShut {NoStop}%
\bibitem [{\citenamefont {Sedrakyan}\ \emph {et~al.}(2015)\citenamefont
  {Sedrakyan}, \citenamefont {Galitski},\ and\ \citenamefont
  {Kamenev}}]{sedrakyan2015statistical}%
  \BibitemOpen
  \bibfield  {author} {\bibinfo {author} {\bibfnamefont {T.~A.}\ \bibnamefont
  {Sedrakyan}}, \bibinfo {author} {\bibfnamefont {V.~M.}\ \bibnamefont
  {Galitski}}, \ and\ \bibinfo {author} {\bibfnamefont {A.}~\bibnamefont
  {Kamenev}},\ }\href {\doibase 10.1103/PhysRevLett.115.195301} {\bibfield
  {journal} {\bibinfo  {journal} {Phys. Rev. Lett.}\ }\textbf {\bibinfo
  {volume} {115}},\ \bibinfo {pages} {195301} (\bibinfo {year}
  {2015})}\BibitemShut {NoStop}%
\end{thebibliography}%

\end{document}